\documentclass[12pt]{iopart}
\renewenvironment{quote}{
   \par\addvspace{0mm}
   \list{}{
     \leftmargin 5mm
     \rightmargin 5mm
     \itshape
   }
   \item\relax
}{\par\addvspace{0mm}\endlist}

\usepackage{graphicx}
\usepackage{dcolumn}

\begin{document}

\title[Hermione and the Secretary]{Hermione and the Secretary: \\How gendered task division in introductory physics labs can disrupt equitable learning}

\author{Danny Doucette, Russell Clark, and Chandralekha Singh}
\address{Department of Physics and Astronomy, University of Pittsburgh, Pittsburgh, Pennsylvania 15260, USA}
\ead{danny.doucette@pitt.edu}

\vspace{10pt}
\begin{indented}
\item{\today}
\end{indented}

\begin{abstract}

Physics labs provide a unique opportunity for students to grow their physics identity and science identity in general since they provide students with an opportunity to tinker with experiments and analyze data in a low-stakes environment. However, it is important to ensure that all students are benefiting from the labs equally and have a positive growth trajectory. Through interviews and reflexive ethnographic observations, we identify and analyze two common modes of work that may disadvantage female students in introductory physics labs. Students who adopt the Secretary archetype are relegated to recording and analyzing data, and thus may miss out on much of the opportunity to grow their physics and science identities by engaging fully in the experimental work. Meanwhile, students in the Hermione archetype shoulder a disproportionate amount of managerial work, and also may not get an adequate opportunity to engage with different aspects of the experimental work that is essential for helping them develop their physics and science identities. We use a physics identity framework to investigate how students under these modes of work may experience stunted growth in their physics and science identity trajectories in their physics lab course. This stunted growth can then perpetuate and reinforce societal stereotypes and biases about who does physics. Our categorization not only gives a vocabulary to discussions about equity in the physics lab, but may also serve as a useful touchstone for those who seek to center equity in efforts to transform physics instruction.

\end{abstract}


\maketitle

\section{Introduction}

\begin{quote}
In lab, I'm usually in charge of writing down the data that we collect, and [my partner] is usually the one doing the physical part.
\end{quote}

\begin{quote}
I think my partners weren't always prepared for the labs, so it fell on me to understand and get the group to finish the lab... I need to be prepared to know what's going on, because they won't.
\end{quote}

Consider the above quotes from students describing their experiences in introductory physics labs. Who do you imagine these students to be? How might students' genders affect the way they experience the traditional introductory physics lab?

The introductory physics lab presents a unique and powerful opportunity for students to grow their physics and science identities. Identity in this sense is the `kind of person'~\cite{GeeIdentity} students consider themselves -- with respect to physics, or with respect to science generally -- and we may understand the lab as contributing to their larger physics or science identity trajectory. Well-designed labs can be particularly effective for identity growth because of their low-stakes nature, which allows students to `tinker' with the apparatus and develop a meaningful and relevant understanding of physics as an experimental science, and because lab-work can be collaborative and engaging for students.

However, as physics lab instruction increasingly adopts pedagogical approaches that include evidence-based active engagement strategies~\cite{BerginProblemBasedCoopLearning,GalanVirtualInquiryLabs,SobhanzadehLaboratorials,DunnettInquiryBehaviors,KirkupPhysLabsEngineers,NeumannLabTransformation,RealTimePhysics,KoenigLabs,AAPTlab} and collaborative learning~\cite{SharmaLabs,ParappillyTeamBased,BerginProblemBasedCoopLearning,DunnettInquiryBehaviors,KortemeyerCollaboration,SavainainenDiscourse,JJSCooperativeLearning, SinghCoconstruction, CochranCollaboration, HellerBook}, concerns have emerged that these types of learning environments might actually increase the `gender gap' even as all students are learning more than they would in traditionally-taught courses~\cite{KarimEBAEgender}. In particular, if physics lab environments are not equitable and inclusive, social interactions around physics may allow for activation of stereotype threats~\cite{KarimStereotypeThreat} and the perpetuation or verbalization of stereotypes about who belongs in physics and who is capable of succeeding in physics~\cite{HazariIdentity}. Additionally, in such an environment, micro-aggressions, discrimination, and harassment~\cite{AycockHarassment} have the potential to stunt the physics and science identity development of students from traditionally-disadvantaged groups if equity is not placed at the center of the learning process in designing the learning environment. 

Likewise, research shows that due to lack of role models and societal stereotypes associated with physics, women in college physics classes report lower levels of self-efficacy~\cite{CauseForAlarm}, are more susceptible to stereotype threats~\cite{KarimStereotypeThreat}, are more often subject to stereotypes related to their competence, and enroll as physics majors and graduate students at markedly lower rates~\cite{AIPgender2019} compared with their male peers. In introductory labs, women average less expert-like responses on E-CLASS~\cite{ECLASSgender}, an assessment of student attitudes toward experimental physics, and may perform different roles when engaging in lab-work with male peers~\cite{HolmesGenderLabs,QuinnGenderLabs}. Research also shows that as they progress in their careers, female graduate students and scientists continue to experience inequities in research labs~\cite{MasculinitiesGonsalves,DanielssonLabs}. Thus, promoting positive physics and science identity development~\cite{AllieIdentityEngEd} by creating an equitable lab learning environment is especially important for students from traditionally-disadvantaged groups, including women, as we seek to rectify longstanding inequities in physics.

This research is concerned with how gender is expressed in an introductory physics lab if no explicit effort is made to create equitable and inclusive learning environments where all students thrive and how it may disadvantage some students. In particular, we analyze our observational and interview data from the lens of students in the introductory physics lab `doing gender'~\cite{MasculinitiesGonsalves, WestDoingGender} while `doing physics'. Thus, if physics is framed, presented, and conducted in ways and in learning environments that are more aligned with traditional conceptions of masculinity and femininity, students are likely to position themselves and perform in response to these conceptions and reconcile `doing physics' with `doing gender'~\cite{MasculinitiesGonsalves,BugFemPhys, SchiebingerFemSci}. In this research, it is in students' navigation of aspects of their gender while doing physics that we seek clues about how we can improve instructional practices, learning environments, and lab cultures to positively impact students' physics and science identity trajectories~\cite{DanielssonLabs}. We note that we recognize that gender is not a binary construct, however, all students in this investigation voluntarily self-reported identifying as male or female.

With this lens in mind, the introductory physics lab is at a crossfire: required for a large portion of the student body in science and engineering, fundamentally collaborative, increasingly adopting active learning approaches,  and largely unattuned to the impact it is having on the physics and science identities of traditionally-disadvantaged students such as women and racial and ethnic minority students. Unlike traditional lecture-style courses, both labs and reformed courses that use collaborative evidence-based active-learning approaches (such as flipped classes) may allow gender stereotypes about physics to become especially salient and relevant. However, physics learning environments should not be allowed to perpetuate negative stereotypes about who can do physics, and about who can develop a strong identity as a physics or science person. Instead, physics learning environments should help all students develop physics and science identities. To that end, this research may provide useful insight for both labs and courses that employ collaborative learning.

The goal of this research was to use reflexive ethnographic observations in introductory labs and individual interviews with students in those labs in order to identify and map out how traditional lab instruction may impact students who work in mixed-gender groups of two or three students in our traditional introductory labs. In these labs, run by graduate student TAs, there is typically no explicit effort made to make the learning environment equitable. We identify two modes of work in which women may be disadvantaged. In the Secretary-Tinkerer mode, men tend to monopolize tinkering with the apparatus while women tend to be found in a note-taking or supportive role. In the Hermione-Slacker mode (Hermione is named for the clever, devoted, hard-working character from the \textit{Harry Potter}~\cite{HermioneLibrarian} series who exemplifies the role in contemporary media), women tend to be thrust into the role of managing the experimental work, communicating with peers and the instructor, preparing for the lab, and doing most of the work in each lab session while their partners make minimal contributions. Finally, we discuss some research-based approaches that may help to reduce the prevalence of these modes of work.

\section{Framework}

We employ an identity framework to analyze how introductory physics lab learning environments affect the development of physics and science identities for female and male students~\cite{CarloneJohnsonIdentity}. In this framework, physics identity pertains to whether students see themselves as a physics ``kind of person''~\cite{GeeIdentity, StrykerIdentity}. We also acknowledge that a student's identity ``is not predetermined and fixed'' \cite{CarloneJohnsonIdentity} and that one's identity is dynamic and ``always being shaped and impacted by one's environment'' \cite{HyaterAdamsIdentity}. An identity framework is ideally-suited to the analysis of students' experiences in culturally-rich settings~\cite{EnrichingGenderPER, GoslingReview} such as the introductory lab because identity framing focuses on and values the experiences of individual students, while avoiding the trap of deficit models that may be interpreted as inadequacies from differences between students. In our case, we seek to understand whether the way that physics lab learning environments are designed ensure that all students develop a stronger identity as a physics or science person.

Three constructs are often discussed in connection with physics identity. Perceived recognition is the degree to which students feel recognized or valued by peers, TAs, instructors, and family as a physics person or a person who is good at physics. Research suggests both that recognition is the strongest influence on the development of physics identity, and that the average perceived recognition by the instructor/teaching assistant in physics courses is larger for men than for women ~\cite{KalenderIdentityGender2}. Interest is a measure of a student's intrinsic valuation of their engagement with physics and enjoyment of this pursuit in a personally meaningful way~\cite{HazariIdentity}. Self-Efficacy (sometimes also referred to as competency belief) is a student's belief in their ability to succeed in a certain situation, task, or domain~\cite{BanduraSEDefinition, BanduraSelfEfficacy}, and may be associated with long-term student persistence~\cite{CauseForAlarm}. The lower self-efficacy may partly be due to pervasive social and cultural stereotypes and biases and the paucity of positive encouragement and support endemic in the field of physics. All three of these factors -- perceived recognition, interest, and self-efficacy -- would, in general, contribute toward the development of a student's identity as a physics and science person~\cite{HazariIdentity,GodwinEngineeringIdentity,KalenderIdentityGender1}.

There are several ways in which the development of a student's physics identity is important in the lab context. A student who develops a favorable and productive identity as a person who is good at physics is likely to engage, enjoy, and learn more in the lab~\cite{DounasFrazerOwnershipOptics}, both during and after the course is finished. Low-stakes tinkering in the physics lab can be an important part of developing interest and self-efficacy in experimental physics and experimental science in general. A student's physics identity is valuable beyond the scope of the introductory physics sequence, even for students who pursue courses of study in which physics may not be directly relevant. For example, physics identity has been shown to be a strong predictor of interest and agency in engineering programs~\cite{GodwinEngineeringIdentity}, and the movement toward competency-based assessments for medical school in some countries~\cite{MCATredesign} makes clear that proficiency and confidence in using physics ideas and scientific ways of thinking are viewed as essential for future doctors.

A variety of prior studies have identified types of interpersonal interactions in labs and similar learning environments that impact student experiences differently according to their gender~\cite{WorkshopPhysicsWomensResponses,GenderRobotsEngineering,HolmesGenderLabs,DayGenderLabs,QuinnGenderLabs}. In one case, women in introductory college physics appreciated hands-on experiences as valuable but expressed frustration about having to adapt to new types of learning in the active engagement work employed in this class \cite{WorkshopPhysicsWomensResponses}. Research on gender in a robotics-based introductory engineering course shows differences in how women and men described experiencing certain learning activities and dealing with challenges, and also suggested that gender differences stem from the competitive aspects of the course \cite{GenderRobotsEngineering}. In physics labs, observational studies have noted that women spend less time tinkering with apparatus\cite{HolmesGenderLabs, DayGenderLabs, QuinnGenderLabs}. 

We may understand why female and male students have different experiences in the same learning context by considering how and why a student may change their behavior in an attempt to align their self-image with societal preconceptions and cultural expectations of what it means to be a male or female physics student~\cite{MasculinitiesGonsalves}. In other words, students will `do gender' while `doing physics' in order to conform to socio-cultural expectations. An identity framework is useful in this case because it provides a means to understand how identity trajectories of students who identify with different genders are shaped by their environment and experiences from the moment they enter the physics labs and how they position themselves and perform differently in the labs~\cite{GoslingReview,EnrichingGenderPER}. 
Our goal, then, is to extend investigations that applied an identity framework to understanding how men and women worked differently in experimental physics research settings~\cite{MasculinitiesGonsalves,DanielssonLabs} by applying this framework to the introductory physics lab setting. 
This can provide guidance for how to improve the introductory physics lab environment to make lab experiences effective for all students, at this crucial time when a positive boost in students' physics and science identity trajectories can set them on a path for growth as physics and science people and mitigate the impact of stereotypes that may otherwise thwart their positive physics and science identity development ~\cite{MasculinitiesGonsalves,DanielssonLabs}.

\section{Methodology}

In order to investigate the introductory physics lab experiences and interactions of students who identify with different genders, we adopted a qualitative, mixed-methods approach that involved ethnographic classroom observations as well as semi-structured interviews with individual students. Both techniques are influenced by the reflexive strand of ethnographic investigation, in which the observer is mindful of their own positioning and background while planning data collection, interacting with participants, and analyzing results. It is through this reflection that blind spots, biases, and confounding preconceptions are identified and accounted for \cite{BuscattoReflexivity}.

Both stages of this work were performed by the first and third authors, each with more than a decade of experience as a physics educator and a variety of personal experiences doing science in different cultures. The former is a graduate student, a White man and former high school physics teacher. The latter is an Asian female physics professor who has taught and conducted PER research since 1995. Throughout this work, these two investigators collaborated extensively to plan, conduct and analyze observations, and shared reflections at weekly meetings and frequently throughout the week as well.

\subsection{Participants}

The participants in this investigation are students enrolled in a stand-alone introductory physics lab at a large research university in the USA. The course is a one-semester introductory lab, which requires the second half of a two-semester introductory physics course as a co-requisite. Two versions of this lab, corresponding to the algebra- and calculus-based physics sequences, are offered. The algebra-based lab is often taken in the third or fourth year of study, and the majority of students who enroll are bio-science majors with an interest in health-related professions. Students in the calculus-based lab are typically engineering or physical science majors, and are more likely to be in their first or second year of study. While the algebra-based sequence is 55\% female and 45\% male, enrolment in the calculus-based sequence is 20\% female and 80\% male. University records at this time do not acknowledge non-binary gender identities.

The labs are run by graduate student teaching assistants (TAs), who are also responsible for most grading in this course. Enrolment is capped at 24 students per lab session. Students are graded for completion of their work and, aside from a post-lab exercise, partners receive the same grade. The introductory physics labs have a reputation for being somewhat easier than other labs typically taken by students in this course such as organic chemistry, introductory biology, or introductory chemistry lab. Students who attend all 12 lab sessions typically receive at least a `B' grade, and most receive an `A' grade.

In both versions, students worked in groups of two (or three, if needed, e.g., if there is an odd number of students or some apparatus is broken so there are less stations available) to complete a thorough and detailed lab procedure during a 3-hour period. Our observations suggest that students self-select into partnerships essentially at random, as they sit down at an open lab bench on their first lab session. The exception is that a very small number of students partner-up before arriving in the lab: we generally see no significant differences in how these partnerships operate. Once formed, groups tend to stay together unless the TA requires a re-shuffling (see Section \ref{implications}). Most students' pseudonyms were chosen by study participants: they reflect the participant's gender but not necessarily their racial or ethnic identity.

\begin{table}
\caption{Participants in this study from the introductory physics lab, along with the pseudonyms of those quoted in this paper.\label{Participants}}
\footnotesize
\begin{tabular}{llll}
\br
Gender & Female & 13 & (Leah, Elisa, Melanie, Bella, Natalie, Paulette, Zara, Liza, Janet, Kamala)\\
 & Male & 5 & (Mark, Lou)\\
Major & Pre-Health Sciences & 12 & (Elisa, Melanie, Mark, Natalie, Zara, Liza, Janet, Kamala)\\
 & Physical Sciences & 5 & (Leah, Lou, Paulette)\\
 & Engineering & 1 & (Bella)\\
Course & Algebra-Based & 12 & (Mark, Elisa, Melanie, Natalie, Zara, Liza, Janet, Kamala)\\
 & Calculus-Based & 6 & (Leah, Lou, Bella, Paulette)\\
 Total & & 18 & \\
\mr
\end{tabular}
\end{table}

\subsection{Ethnographic Observations}

The experiences that affect students' identity trajectories can be subtle and hard to identify. External observers, however, may be better-positioned to see how words, body positioning, and the manipulation of physical objects can contribute to student's experiences in the lab. We conducted observations many times over the course of the semester. These observations targeted six introductory lab sections during each of the fall 2018 and spring 2019 semesters. Each of the twelve sections was run by a different graduate student TA, who was informed in advance of the observation and asked to briefly introduce the observer at the start of the lab session. Observations lasted at least 1 hour each, in order to develop a fuller understanding of the student interactions that were being observed. In total, more than 100 hours of such observations were completed.

We took on the role of non-participant observers~\cite{OteroGettingStarted}. During our observation sessions, we sat on a side-bench of the laboratory and observed the students and TA while taking notes of what we saw and heard, as well as our reflections on what they might mean. An informal observation protocol~\cite{OteroGettingStarted} was adopted, and iteratively refined, as we sought to understand factors that might affect students' identity trajectories in the lab. With practice, and after comparison of notes between observers, we came to identify particular items of interest: comparing same-gender with mixed-gender groups, the work done by students in mixed-gender groups, and the nature of the students' discussions about their lab-work.

In line with our reflexive approach to investigation, we sought to fulfill three goals in how we positioned ourselves during our observations: acceptance, detachment, and reflexivity~\cite{BuscattoReflexivity}. First, we aimed to position ourselves in such a way as to not influence the normal behavior of the TA or students. Sitting at the side of the lab helped in this effort, but we also engaged in a small amount of discussion with a few students (offering brief advice on the apparatus, asking Socratic questions about concepts, etc.) to establish the idea that we were friendly and unobtrusive. This was largely successful for the students, who were typically focused on their lab-work and ignored the observers. In follow-up discussions, some of the observed TAs agreed that our presence did not noticeably affect what the students did in the lab.

Our second goal was to keep sufficient distance between the lab participants and ourselves in order to make balanced observations. To this end, we kept discussion with students and TAs to a minimum (less than 10\% of observation time). The third goal, reflexivity, required continuous reflection on how our own backgrounds and preconceptions may affect what we see, and what we deem important. To achieve this, we sought to pay attention to each individual student, to take their lab experiences at face value, and to discuss as observers these issues in order to come to a shared research agreement. 20 hours of our observations were done simultaneously, allowing us to compare notes and confirm that we observed similar events, behaviors, and interactions. In reviewing our detailed and thorough notes, we are confident that we were successful at maintaining suitable detachment and reflexivity in our observations.

\subsection{Semi-Structured Interviews}

Based on our classroom observations, we identified students whose perspectives and experiences we thought would (a) provide a cross-section of the students who enroll in the lab classes, and (b) had experiences and perspectives that would be valuable for us in understanding student interactions in the lab. These students were invited to participate in hour-long interviews, for which they were compensated with a \$25 payment card. Roughly half of the students who were invited agreed to be interviewed and we conducted a total of 18 interviews at the end of the fall and spring semesters during the 2018/19 academic year.

Our reflexive ethnographic observations suggested differential gender effects with negative impacts on women, so we aimed for an interview pool that included more women's voices. Our decision was supported by the fact that only two of the five men we interviewed were aware of these effects (perhaps experiencing a blindspot~\cite{BanajiBlindspot}) while all of the women were able to describe at least one way in which men and women experienced the lab differently. In addition, we sought particularly to speak with students from mixed-gender partnerships, as these seemed to be the locus of gendered inequity of opportunities, based on the ethnographic observations. By comparison, we observed that students who worked in same-gender groups tended to collaborate much more effectively and equitably. Of the 18 participants who agreed to participate in interviews, 13 identified as female and 5 as male, and all but one described working in a mixed-gender group for at least part of the lab course (we note that most students in the lab course stayed with the same partner throughout and only a few occasionally switched). All 18 participants worked in groups that were stable over the course of the 14-week semester.

Drawing on our observations we assembled and refined a list of potential interview questions to serve as our interview protocol~\cite{OteroGettingStarted}. These included questions about the student's background and prior lab experiences; interactions with other students and the TA; thoughts on the structure, mechanisms, and effectiveness of the course; and experiences with task division, including gendered division of labor. Despite the long list of questions, we sought to make these semi-structured interviews conversational in nature to give participants the opportunity to express themselves freely, dig deeply on critical issues, and remain comfortable and safe. The investigators used the list of questions to gently steer the conversation in the directions specified by the interview protocol. Most participants required little prompting and were keen to share openly. All interviews were audio-recorded and transcribed.

\section{Results and Discussion}

\subsection{The Secretary and the Tinkerer}

Students in the physics lab have a wide variety of background experiences. Some have taken AP Physics in high school, while others went to schools that didn't offer it. Leah, a high-achieving chemistry major, described why she didn't take physics in high school:

\begin{quote}
I had never had physics in high school at all. My school pushed for biology and chemistry for girls, and physics for guys... So when I came here I had no clue about anything about physics... I was clueless in a sense about physics. Physics I and II, the calculus-based ones, were [a] little fast for me but a good speed for everyone [else]. The physics lab seemed a lot slower paced, so it was really good for me but it was kind of boring for other people that were very, very, very good at physics...
\end{quote}

Leah's high school experiences established a clear picture of who can be a physics person, so it is unsurprising that she expressed a low level of physics self-efficacy and did not see herself as a person who can be good at physics. Furthermore, her prior preparation meant that when she got to college, Leah had little confidence in her ability to do physics. Her low self-efficacy is clear when she compared herself with peers, whom she perceived to be mastering physics concepts much more quickly than she was. However, Leah acknowledged that when she compared her grades with those of her more-confident classmates, she saw that she was doing just as well as them. 

\begin{quote}
There would be times when I would feel like I am not good at physics, I am not good at it. But we would get tests back... I was very comparable to them, but I still felt like, `Oh, it's not my thing, I'm not very good at it.' But here I am, and they think they are very good at it and I'm doing just as well as them.
\end{quote}

While Leah was certainly doing well in class, her physics identity was stagnant because her low self-efficacy prevented her from internalizing the idea that she was developing mastery of physics concepts. Even though Leah was telling her (male) peers that they were ``very good at physics'', no-one was communicating that type of message to her or recognizing her success. This conflict is typical for women enrolled in introductory physics and even though men and women perform equally in introductory physics at the institution where this study was carried out, men report substantially higher self-efficacy~\cite{WhitcombSEandGradesPERC}. In negotiating a role in the physics lab, it is Leah's low self-efficacy as a physicist -- developed through the lack of support at high school, the encouragement she didn't receive as a student, and a shortfall of recognition when she did begin to demonstrate mastery -- that may have led her to adopt a secretarial role.

Mark, a microbiology major in his final semester, also hadn't taken physics in high school. However, unlike Leah, Mark was given opportunities to play with circuits as a child, and to learn how to work with electronics through school research programs. These prior experiences helped reinforce Mark's interest and self-efficacy in science.

\begin{quote}
I've taken apart a lot of things. I've done work with Arduinos, kind of, building my own circuits... [My father is] a chemical engineer, so I always had something I could work with when I would take things apart, until I bought my own things... some of [the Arduino work] I had done with my research experience outside of school, having to design some things, measuring bacteria and things like that. One of them, I did this summer program where we build a little thing to switch LEDs off and on, and also to measure absorbance inside cultures.
\end{quote}

Mark's prior experience led him to adopt the role of the Tinkerer in his lab group. He recognized that this meant an unequal division of labor in his group. When asked explicitly about why male students sometimes took over the apparatus in the introductory lab, and what could be done about it, Mark replied:

\begin{quote}
I would say maybe some of the labs that had a more technical set-up, I would do more of that. And then while I was setting that up, she would be waiting... I'm usually in front of the machine so I'm usually handling that while she's inputting all the data. And that's maybe something to think about, maybe changing the roles.
\end{quote}

We see here an example of masculine lab behavior being replicated along gender lines~\cite{MasculinitiesGonsalves}, to the benefit of Mark at the expense of his partner, Elisa. Furthermore, Mark attempted to blame his dominance of the apparatus on his seating location. We found this attribution by male students when questioned to be common, but spurious, as we observed most students alternate locations readily as they do their lab work. 

Elisa agreed with Mark's description of the unequal task division in their partnership, but speculated he must have taken advanced physics classes to have such a high self-efficacy with the lab apparatus (in fact, he had neither taken high school physics, nor had he taken any physics classes she had not). Here, again, notice how asymmetric engagement with the lab-work only provided opportunities to Mark, potentially bolstering his physics identity development while hindering Elisa's. By assuming that he must have taken advanced classes and allowing him to do the tinkering, Elisa appears to recognize Mark's practical skills and self-confidence. This is a message that may have bolstered his self-efficacy, even though he hadn't actually taken such classes.

On the other hand, Elisa was doing the other work: she didn't get to develop expertise with experimental techniques in a low-stakes environment, didn't get acknowledged by her partner positively, and therefore didn't get an opportunity to develop her physics and science identities. The types of task they each performed and the opportunities she and her partner had in the physics lab are likely to further increase the gap between their self efficacies when it comes to tinkering and the associated learning in the lab. Elisa elaborated, describing a typical day in the lab as follows:

\begin{quote}
He liked to do a lot of the setting-up and he knew what was going on, more than I did. I felt like we both tried to split [it] up, so it wasn't one person doing all the work. I like to do the data entry and stuff, so often I would do that.
\end{quote}

This division of work into Tinkerer and Secretary roles was a theme we saw repeated frequently, in both algebra- and calculus-based labs, when students worked in mixed-gender groups. In most cases, the Tinkerer tended to be male, and the Secretary tended to be female. When the Secretary-Tinkerer split happened, as with Elisa and Mark, students typically thought of it as a fair division of labor. Melanie, a biology major, described how she and her partner split the work:

\begin{quote}
In lab, I'm usually in charge of writing down the data that we collect, and he's usually the one doing the physical part.
\end{quote}

While the Secretary-Tinkerer task division looks fair on its surface, there are two big reasons why it can be a deleterious approach to work. First, this division can reinforce a power imbalance in team-work that deprives the Secretary of the opportunity to be a scientific investigator. Lou, the partner of Leah (above) and a fellow chemistry major, described a moment when he interfered with his partner's attempts to contribute to building a complex circuit:

\begin{quote}
Sometimes I get a little carried away with getting things to work. If Leah would come over and try to change things, I'd be like, ``I've almost got it.'' That's just my personality. 
\end{quote}

Leah described the same type of interaction in her interview. Traditional gender roles were being enacted here: the man as authoritative, and the woman as responsive. However, Leah wanted to do her fair share of the tinkering and recognized the inequality in their division of the work. The following situation was a rare case of the Secretary being willing to speak up and risk conflict, and may be seen as arising from a mismatch between Leah's relatively high level of initiative as a learner and the expectation that Secretaries have a more passive role.

\begin{quote}
In the circuits labs, he kind of took over the experiment... the next week, I was kind of like, ``okay, give me that wire.'' I tried to do more of the trying to plug in and see what's going on.
\end{quote}

The second reason the Secretary-Tinkerer split is deleterious is that it deprives both members of practice with the other type of working. Since the physics lab is often the only place students learn to do hands-on experimental physics, the Secretary stands to lose more from this task division than does the Tinkerer. Many of the skills recommended by the AAPT~\cite{AAPTlab} such as constructing and using apparatus, making measurements, and troubleshooting problems cannot be learned by watching a partner. As a contrast to her introductory physics lab, Bella described a digital circuits lab she took as an engineering student:

\begin{quote}
It's mainly the guys who are building the labs. And the women are mainly having to figure out the software and the calculations... I don't know, maybe it's a perception that men are better at things that require the use of hands?
\end{quote}

When asked whether the gender split deprived women of opportunities to learn, Bella explained that she felt under-prepared for a mid-term practical assessment in her engineering lab:

\begin{quote}
Definitely! It definitely does. On the practicum, I remember thinking, `Dang, my partner always did this part of the lab.'
\end{quote}

Although most of the interview participants discussed short-term impacts, the Secretary-Tinkerer split can also have long-term negative consequences. In particular, this inequitable task division deprives women of the opportunity to tinker in a low-stakes environment, which is necessary for developing one's physics and science identity as a person who can handle the equipment and experiment.

\subsection{Hermione and the Slacker}

While the Secretary-Tinkerer mode of work deprives female students of the opportunity to tinker with apparatus, which is a critical part of the lab and essential for identity development as a physics or science person, we observe a very different effect in a second mode of work that is equally salient. In this case, a student, typically female, ends up shouldering a disproportionate amount of the work and compensating for the shortcomings of their partner(s). Such students take on the responsibility of ensuring the work gets done when their partners fall short, but are more than just a project manager. In the physics lab, Bella described working with two partners and asking one of them a question, only for him to turn the question back on her because he hadn't prepared for the lab and did not want to think about it.

\begin{quote}
I feel like I did a lot of the thinking for the group... [When I asked him a question] he would be like, what do you think?
\end{quote}

Typically, students who adopt this Hermione archetype see it as necessary in order to complete their lab-work because their partner, the Slacker, appears to be uninterested. Like the Secretary-Tinkerer split, the Hermione-Slacker task division is one that seems to strengthen as partners work together for more than one lab session, as the partners recognize that the other person would be willing to pick up the slack.

The Hermione-Slacker mode of work seems to be especially prevalent in groups of three students, although we also observe it in pairs. It may partly be that the student(s) realize that their lab partner will make sure things get done and, thus, they will receive a good grade with minimal effort. Natalie explained her disappointment that her partner wasn't contributing:

\begin{quote}
I like being on a team... Seeing that he puts in as much effort as I put in... Because I don't see that effort coming from him, I've had to step up to make up for that effort so we get it done with.
\end{quote}

The lack of engagement or initiative from Natalie's partner, however, went beyond merely not contributing. She described how her partner's disinclination to participate led to her skipping a portion of the lab report that was not explicitly graded:

\begin{quote}
In the beginning of the semester, I would try to do the analysis questions just because I wanted to understand it more, and he was like, we don't have to do this, there's no reason to doing this. So I kind of gave up on that portion.
\end{quote}

As a result of this partnership, Natalie's opportunity to grow her expertise and interest in physics was stymied, and in the rest of the interview it was evident that subscribing to her partner's lackadaisical approach to doing the lab just to get a grade may have negatively impacted her physics and science identity development.

Despite being a physics major, Paulette's male partner seemed to have little interest in completing the lab, let alone contributing equally to the mental and physical labor required to complete the work. This put her in the awkward position of needing to repeatedly ask him to contribute to work for which he was receiving a grade and, perhaps worse, forced Paulette into a traditional -- almost maternal -- role, depriving her of the opportunity to dig deeper and develop her self-efficacy as a subject-matter expert.

\begin{quote}
Well, my partner's a little lazy... Sometimes he's on his phone and stuff, and I'm just like, `get off your phone.' He helps when I ask. I'll be like, `hey, can you do this?' But he doesn't really start doing stuff himself most of the time. I'm like, `I'm not your mom.'
\end{quote}

As time went on, Paulette explained, he took increasingly-long and increasingly-frequent breaks from the lab, and contributed less and less to the lab-work they should have been sharing equally. She described asking him to help, but he was so detached from the entire task that he would not even know where they were in the lab procedure or what needed to be done.

We observed Hermiones taking on a variety of tasks, including preparing for lab when their partners did not, managing the work-flow, assigning small tasks to their partners and monitoring their progress, communicating with the lab TA and other groups, and ensuring the data collection was complete before leaving the lab room. We also saw Hermiones take on the labor of reconciling different and sometimes conflicting instructions, methods, and conceptual ideas. It added up to a lot of commitment and effort, and so frustration with a partner's lack of preparation is a common theme for students such as Zara in this role. Here, she described what it was like when her partners didn't adequately prepare for the lab, and her experience the one week the group had to stay late in order to finish their work because she wasn't as well prepared.

\begin{quote}
There was one lab where, working with circuits... that was very difficult for me. Maybe it's just because during the week I didn't have a very good week or something. I really struggled understanding it. I think my partners weren't always prepared for the labs, so it fell on me to understand and get the group to finish the lab... I need to be prepared to know what's going on, because they won't.
\end{quote}

Despite the disproportional amount of time and effort she invested into the lab, Zara either didn't receive or didn't internalize recognition from her peers. When asked if she was the expert in her group, she laughed and said:

\begin{quote}
I definitely would not call myself an expert. Maybe I read the lab manual more?
\end{quote}

According to the identity framework, perceived recognition should stimulate development of Zara's physics identity. However, because her lab participation was managerial, rather than focused on the physics or hands-on parts of the lab work, the recognition she received from her partner was -- in her view -- related to the project management, rather than mastery of physics concepts and skills. Moreover, it appeared that Zara wasn't internalizing the little recognition she did receive from her peers, and so she appears to have experienced little identity development as a physics person.

Like Zara, Liza described her Hermione role in a way that situated her as doing necessary work to accommodate an unprepared peer:

\begin{quote}
He didn't read the manual every week, a lot of the time it was me telling him what to do... do this, do this, and it would be me doing the note-taking... I felt like I was controlling from that position.
\end{quote}

The Hermione archetype can disadvantage students who adopt it in part because they do the majority of the work while receiving the same learning experience and/or grade. Even worse, the managerial work they do takes them away from the tinkering and sense-making activities that could help them to develop their identities as physics and science people. Janet described spending a large portion of her time mediating between her partner and the TA, asking questions to the TA about things she already understood, in order to appease her partner after he hadn't bothered with the pre-lab reading and expressed doubts about her explanations of the tasks they needed to do. 

\begin{quote}
It's like, you're wasting my time because you're unprepared. Well, now I'm not able to learn as well because I'm spending so much time asking [his] questions [to the TA] that I don't really need to ask, because I know what's going on. It's wasting my time...
\end{quote}

Since Hermione-role students are typically situated as the hard-working one in their partnerships, these students tend to attribute their successes to their exertions rather than their physics competence, which could again shortchange their physics and science self-efficacy and identity development. And because they need to be so laser-focused on getting everything done for their group, there is little time or capacity for Hermiones to develop higher levels of self-efficacy and interest in physics, and to grow their physics and science identities, through their experimental work.

Kamala, a high-achieving pre-med student who managed a group of three, praised the skills of one of her partners:

\begin{quote}
He's very good at equipment, so even if he doesn't necessarily read the lab, he's just one of those people that has very good problem-solving skills when there's hands-on things.
\end{quote}

On the other hand, when it came to her own expertise, she rebuffed credit from her partners, interpreting what they say as not genuine, saying:

\begin{quote}
They have an impression that I'm just better at physics than they are. Or I'm just smarter at this stuff than they are. Which isn't necessarily true. It just comes down to... are you willing to push the group forward in terms of knowing what the next thing to do is?
\end{quote}

In effect, then, Kamala praised her partner for practical work, which he did because of his confidence with the equipment but without reading the lab manual, while she appears to have internalized no recognition for her mastery of the physics concepts or experimental procedures. In part, this was because she felt she was essentially managing the lab work for her group in order to make sure it got done. This is a common theme in these interviews: women displayed lower self-efficacy than men, and were more likely to attribute their success to external factors such as hard work rather than to their own developing mastery of experimental physics. By focusing on managerial work, women who adopted the Hermione archetype received recognition that was either not relevant to their physics and science identities or that was interpreted as not being genuine. They consequently appear to have experienced physics and science identity growth that was stunted in comparison with their peers in same-gender groups, or in comparison with the men in the class who adopted Tinkerer roles.

\section{General Discussion}

While the Secretary-Tinkerer mode of work has been documented before in research on STEM education \cite{BarthelemySexismGradSchool, KrethWritingEngineering, CamachoMicroagressions}, here we introduce the Hermione-Slacker mode for the first time. We believe that this taxonomy will help educators conceptualize and reflect on the ways in which students may be disadvantaged by gendered modes of work in the physics lab and other places in which students are doing science together. These archetypes are both salient and ubiquitous in mixed-gender groups, especially when compared with same-gender groups.

Applying the identity framework, we find that both Secretary and Hermione archetypes can act to stunt the development of physics and science identity for women in these roles. Women in secretarial roles, like Leah, Elisa, and Melanie, are denied the opportunity to actively engage with the apparatus in a low-stakes environment of the lab, and thus do not benefit from this opportunity to grow their interest in experimental science. Thus, it is unsurprising that Secretaries typically describe a transactional view of their lab-work: they do what is required, and do not see themselves as undergoing growth in their identity as physics or science science people as a result of the lab course.

In the same way, discussions with women in the Hermione role, like Natalie, Paulette, Zara, and Kamala, suggest little growth in their physics and science identities as a result of this physics lab. Because they are pushed to adopt a managerial (or even maternal) role, they see themselves primarily involved in getting things done, leaving little time to deeply engage with work that might stimulate growth in their interest and self-efficacy in physics, or the consequent development of their identities as physics and science people. And while their partners sometimes recognize them for their leadership in the introductory physics lab, they rarely appear to internalize those types of recognition for their accomplishments in terms of being good at physics.  In total, of approximately 20 students in the Hermione role we identified during the observation phase of our work, none of them were men. 

Overall, then, we find that students whose negotiation of ``doing gender'' and ``doing physics'' results in them adopting Secretary and Hermione roles experience interactions in the lab that limit their physics and science identity development. Returning to the identity framework, we note that students who adopt the Hermione or Secretary roles receive inadequate recognition as scientists. Hermiones, in particular, may receive recognition from peers that is either not perceived as genuine or inadequate compared to the work they do. Likewise, the task division encountered by students in both these roles may have provided fewer opportunities to develop an interest in experimental science, but this was less explicit and salient in the interviews. Finally, students in the Hermione or Secretary roles spend time on managerial or notekeeping work that doesn't promote development of their self-efficacy as scientists. Since these two roles tend to be occupied primarily by women, given the existence of pervasive societal stereotypes about physics that can disadvantage women, this issue deserves careful attention.

We emphasize that this analysis is focused on the impacts of gendered roles in the lab on students.  For example, many students in Tinkerer-Secretary partnership may have good intentions.  In particular, some students who adopt a Tinkerer role may view themselves as doing extra work to the benefit of their partner, while their Secretary partners may believe that stepping back from the apparatus allows them both to finish the lab efficiently. 

\begin{figure*}
\begin{tabular*}{\textwidth}{c @{\extracolsep{\fill}} ccc}
\includegraphics[width=.25\textwidth]{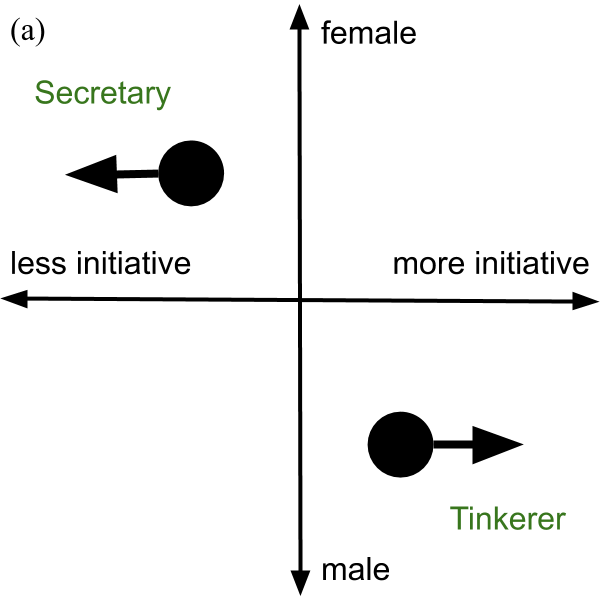} &
\includegraphics[width=.25\textwidth]{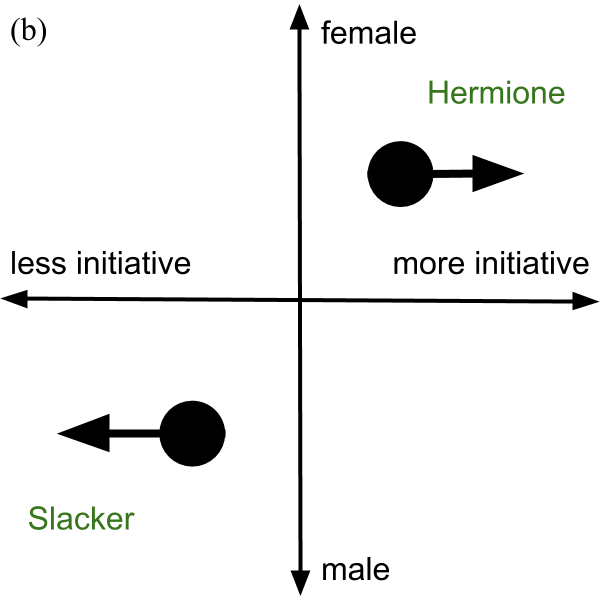} &
\includegraphics[width=.25\textwidth]{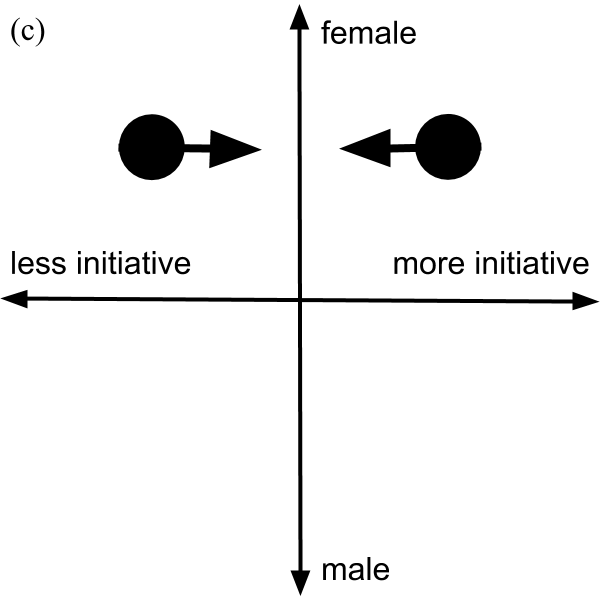} \\ \\
\includegraphics[width=0.25\textwidth]{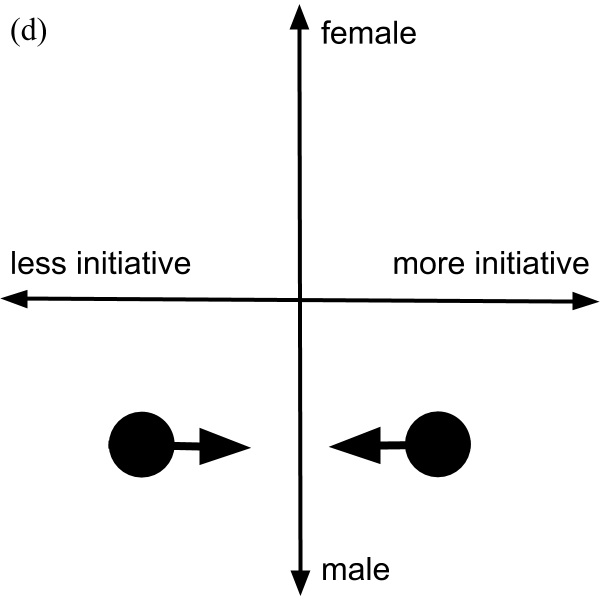} &
\includegraphics[width=0.25\textwidth]{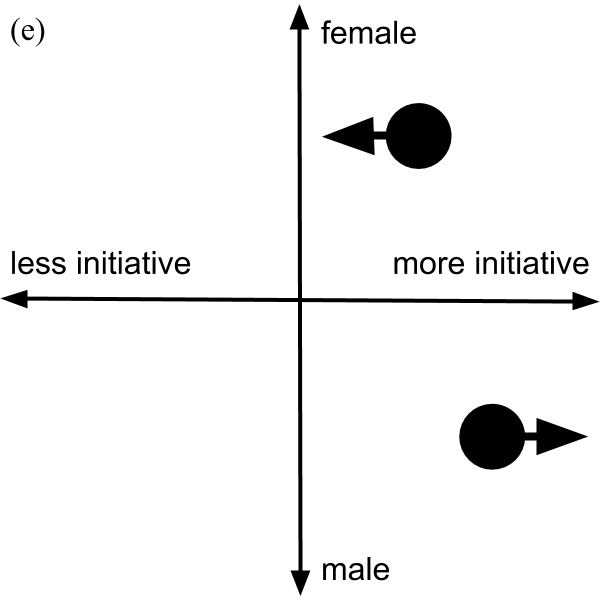} &
\includegraphics[width=0.25\textwidth]{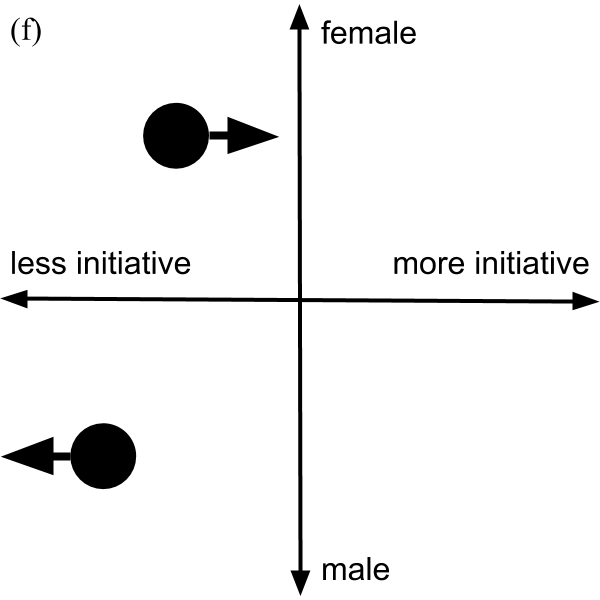} \\
\end{tabular*}
\caption{A proposed model to account for how female and male students settle into adopted gendered modes of work in mixed-gender groups.\label{GenderInitiative}}

\end{figure*}

Some of the classroom observations that two of the authors conducted were at the beginning of the semester (first lab class) to observe how students selected their partners and settled into different roles. Based upon these observations for mixed-gender and same-gender groups, we propose a model for how the gendered-roles solidified in many mixed-gender groups compared to the same-gender groups. 

We present in Fig.~\ref{GenderInitiative} one possible way to visualize the dynamics we identified during our observations that were corroborated by interviews. In this model, a student's initiative - their willingness to do work - in the lab is plotted horizontally, while the vertical axis shows a student's gender. Fig.~\ref{GenderInitiative}a shows the typical dynamics we observe when a woman with lower initiative begins to work with a high-initiative man: he tends increasingly to take over the experiment, adopting the Tinkerer role, and she tends more toward the Secretary role. Similarly, Fig.~\ref{GenderInitiative}b shows what we typically observe when a high-initiative woman begins to work with a low-initiative man: she adopts a Hermione role, and he becomes a Slacker. We observe this type of dynamics that drives this task division throughout the lab period, but they are especially pronounced during the first hour that a pair of students is beginning to work together.

Our observations suggest that unlike in mixed-gender groups, the symmetry breaking and ``phase separation'' into different roles generally does not seem to occur in same-gender groups. In fact, in our observations, the general contrast between the mixed-gender and same-gender groups in this regard was striking. As depicted in Fig.~\ref{GenderInitiative}c and Fig.~\ref{GenderInitiative}d, typically, two students of the same gender who work together - regardless of their initial differences in initiative - tend to achieve an equilibrium, adopting similar types and amounts of work. In these same-gender partnerships, there is little psychological distance~\cite{PsychologicalDistance} (a measure of the similarity between two people based on their characteristics, their behaviors, and the social groups to which they identify) between the partners. This may be a relevant factor in determining whether two students will collaborate effectively.

We also noticed, in our observations, a few cases of mixed-gender groups that began the lab period with comparable initiative. This was the case for Leah and Lou, as described in the previous section. Even though they both started off with a high level of initiative, the female student appeared to display slightly less initiative (which may be due to her gender identification in the mixed gender group), and this small difference was exacerbated by the collaboration, as shown in Fig.~\ref{GenderInitiative}e. However, Leah's determination to take an active role in the lab-work meant that neither she nor Lou moved very far on the diagram, but also introduced tension to their interactions sometimes as discussed earlier. We find that when there was tension in groups, it typically came from conflict between students' desired form of participation in the lab and the role-division described here. In another group, we observed two lower-initiative students (compared to the average of the class) struggling to complete the lab-work until the student with slightly-more initiative started to put in more effort. In this case, shown in Fig.~\ref{GenderInitiative}f, the female student appeared to have slightly more initiative originally, and was pushed toward adopting a Hermione role while her partner became more of a Slacker.

In summary, we propose a preliminary model of lab dynamics in which gender identification of students, as a type of psychological distance, acts to push similar (e.g., same gender) students toward fair and equitable work-patterns, while driving dissimilar (e.g., different gender) students toward the inequitable archetypes described above. Future work is needed in order to refine this proposed model and understand how labs in which students partner with each other may produce the type of dynamics we observed. This may include looking at constraints and affordances of the lab such as how the grades are assigned and the impact on students' physics and science identities of emphasizing the long term utility and value of dividing all aspects of the lab equitably. Furthermore, while this paper has emphasized gendered instances of these archetypes, we note that task division in our labs can also be influenced by our students' racial and ethnic identities. Unpacking and understanding this effect would require an intersectional lens, and is beyond the scope of this paper.

\section{Implications for Practice\label{implications}}

A key question is how to address these inequitable modes of task division. Below, we describe five approaches we have started to implement in our labs, which appear to be promising. In our observations, these approaches seem to be beneficial for students from all four archetypes identified above. Just like Secretaries and Hermiones, Tinkerers and Slackers benefit from increased accountability (including grade incentives), more clearly defined responsibilities, and opportunities to renegotiate their role in group work. 

First, regularly changing group composition may help to reduce some types of inequities in group-work~\cite{HellerBook}. When students work together over several weeks, we see that their adoption of inequitable modes of work (including task division) becomes solidified over time. In labs that changed groups mid-semester, our observations suggested more-equitable work in the second half of the semester compared with the first half of the semester.

A second, often-recommended, approach is to assign (and rotate) roles within student groups \cite{HellerGroupwork}. Recently, we have observed some success in reducing the occurrence of some inequitable task division in labs where the TA (after being prepared to do so via professional development) required that student partnerships take on the roles of `experimenter' and `recorder', and alternate weekly. Students were generally willing to play along and stick to their roles, but we noticed that some recorders would take over parts of the experimentation role if their partners struggled. Likewise, students who were recorders during a given lab session had no opportunity to develop practical skills with the apparatus for that particular lab, and so even though they might get as much experience with the apparatus as their partner, neither of them gets as much experience as they would both get if they collaborated equitably. A series of `checkpoints' provided a grade incentive to students to ensure they fulfilled their roles.

Third, we found that isolated minorities -- such as a woman in a group of three with two men -- were particularly vulnerable to the archetypes described above. Thus, we endorse the advice \cite{HellerGroupwork} to avoid isolating minorities if possible. However, this idea can be usefully extended by considering what happens when a minority student is in a class without any peers from their minority group. During our interviews, one student described her experience as the only female student in an honors physics lab. Despite being friends with many of her male peers, none were willing to partner with her for investigations that were too comprehensive to effectively complete alone. She eventually dropped out of that course and enrolled in the regular (non-honors) lab the next semester, but told us that she would have stuck with it if she had had a fellow female student with whom she could have worked. Thus, we suggest that instructors be careful not only to avoid isolating underrepresented minorities in groups of non-minority students, but also to take care not to allow underrepresented minority students to be isolated without the social resources they need to complete their work with the same level of support as other students.

Fourth, we note that a small amount of previous experience can provide a big boost to a student's self-efficacy when it comes to lab-work at this level. Mark and Lou attributed their tinkering predispositions to extracurricular science activities they experienced at school, for example. One approach, then, is to ensure that all students have an opportunity to tinker unimpeded during the first few minutes of the lab session. Since women are less likely to have had such experiences due to societal biases and stereotypes~\cite{CrowleyGirlsMuseum}, perhaps the lab room could be equipped with enough apparatus for each student to build a few simple circuits, use a caliper, or set up a lens individually before undertaking the cooperative part of the experimental work. In this way, individual `tinkering time' could be built into the lab-work, and students could be coached to do it alone, and not to interrupt their partners' preliminary tinkering.

Finally, recognizing that collaboration is a skill like any other, we have begun to develop and systematically evaluate lab tasks that explicitly divide the learning tasks between partners. For example, Student A is assigned to develop the theoretical prediction while Student B carries out the measurement before they share, and switch roles for the next part of the experiment. This structured work can act as a scaffold over the first few weeks of the semester, and can then be slowly withdrawn as students become more familiar with the expectations for equitable collaboration in the physics lab and more capable of working in an equitable way over the course of the semester. Here, too, we adopted a mixed grading scheme that partially accounted for individual contributions to a group lab report.

As a further issue, introductory physics labs at large research universities are often run by graduate teaching assistants. In such cases, professional development to establish `buy-in'~\cite{GoertzenTABuyin, WilcoxTABuyin} for the principle of equitable learning, designing approaches for teaching assistants to use in their labs, and instructing and monitoring the use of these approaches will be an essential consideration for such settings~\cite{DoucetteLabTAsPERC}. In labs run by our graduate student TAs, we see little or no impact from any of the above strategies when the TAs do not believe in their necessity and benefit.

While the adoption of the archetypes described and illustrated in this paper -- the Tinkerer, the Secretary, the Slacker, and Hermione -- is symptomatic of inequitable learning in the physics lab, these archetypes also serve as a way to understand the nature of the inequities. It is our hope that these labels provide a vocabulary for discussing equity in the lab and a reference point as we work to transform introductory labs into places where all students develop positive identity as physics and science people.

\ack{}
We wish to thank the anonymous students who served as our research partners for their time and insight; Natasha Holmes, Kathy Harper and Bob Devaty for insightful discussions; and the NSF for grant PHY-1524575.

\section*{References}
\bibliographystyle{iopart-num}
\bibliography{the}

\end{document}